\documentclass[12pt,a4paper]{article}
\usepackage[margin=1in]{geometry}
\usepackage{amsmath,amssymb,bm}
\usepackage{microtype}
\usepackage{cite}
\usepackage{hyperref}

\title{\textbf{Quantum Uncertainty and Particle Trajectories:
From Heisenberg and Schr\"odinger to Constructive Dynamical Models}}
\author{Partha Ghose\thanks{Email: partha.ghose@gmail.com}\\
Tagore Centre for Natural Sciences and Philosophy,\\
Rabindra Tirtha, New Town, Kolkata 700156, India}
\date{}

\begin{document}
\maketitle
\begin{abstract}
Quantum uncertainty emerged from two strikingly different routes to quantum
mechanics. Heisenberg's matrix mechanics replaced the classical description
of atomic orbits by relations among observable transition quantities and led,
in 1927, to the uncertainty relation and its influential measurement-based
interpretation. Schr\"odinger's wave mechanics grew instead from the
Bose--Einstein and de Broglie developments and restored a continuous
space-time description through matter waves. The two formalisms were soon
shown to be closely related, but neither historical route by itself settles
the ontological question of whether particles may possess underlying
trajectories.

After reviewing these developments, including Einstein's criticism of a
simple observability doctrine and Mott's wave-mechanical account of straight
cloud-chamber tracks, we distinguish precise quantum uncertainty relations
from the stronger claim that quantum mechanics excludes definite microscopic
histories. We then compare Feynman paths, Bohmian trajectories, Nelson's
stochastic mechanics, and finite-speed persistent Goldstein--Kac dynamics.
The Kac process has continuous, piecewise differentiable finite-speed paths;
its diffusion limit approaches Wiener kinematics, while analytic continuation
of its two-sector dynamics is related to Dirac evolution. These examples show
that quantum uncertainty relations constrain observable statistics without,
by themselves, uniquely determining an underlying path geometry.
\end{abstract}


\section*{Introduction}

The uncertainty principle is one of the most celebrated consequences of
quantum mechanics, but its historical origin is inseparable from the two
very different routes by which modern quantum theory was created. In
Heisenberg's route the decisive step was to abandon the attempt to describe
unobservable electron orbits inside atoms and to formulate the theory in
terms of quantities connected with observable transitions. In
Schr\"odinger's route, by contrast, the central idea was to associate wave
phenomena with material particles and to formulate quantum mechanics as a
wave equation in configuration space. The subsequent mathematical
relationship between matrix and wave mechanics did not erase this
difference in physical motivation.

This history is directly relevant to the modern interpretation of quantum
uncertainty. The exact uncertainty relations are mathematical statements
about dispersions of quantum observables. The stronger assertion that a
quantum particle cannot possess a definite microscopic history is an
additional interpretive claim. The aim of the present paper is therefore
twofold: first, to recall concisely how the question of observability,
waves, and trajectories entered the foundations of quantum mechanics; and
second, to examine whether the uncertainty relations themselves rule out
underlying trajectories. The discussion is organized around the two
historical routes and a third, constructive part devoted to possible
trajectory descriptions.

\section{Heisenberg's Approach: Observables, Matrix Mechanics, and Uncertainty}
\label{sec:heisenberg_history}

\subsection{From atomic orbits to observable quantities}
\label{sec:heisenberg_observables}

The old quantum theory retained the classical picture of an electron moving
on a definite orbit, while imposing quantum conditions on the allowed
motions.  By the early 1920s this hybrid picture had become increasingly
difficult to sustain.  Atomic spectroscopy provided experimentally accessible
transition frequencies and intensities, but it gave no direct access to the
supposed trajectory of an electron between emission or absorption events.
Heisenberg therefore adopted the principle that a physical theory should be
formulated in terms of quantities that were, at least in principle,
observable.  In his 1925 paper he accordingly abandoned the classical
electron orbit and sought to construct quantum mechanics entirely from
quantities associated with transitions between stationary states
\cite{Heisenberg1925,JanssenDuncan2023}.

Heisenberg later related this attitude to the lesson he believed he had
learned from Einstein's analysis of simultaneity in special relativity:
concepts without an operational connection to possible observations should
not enter the fundamental formulation of a physical theory.  The electron
orbit inside an atom appeared to him to be just such a concept.  Frequencies
and intensities of spectral lines were observable; the supposed orbit that
produced them was not.

The limitations of this reasoning were brought sharply to Heisenberg's
attention in a remarkable conversation with Einstein following a lecture
Heisenberg gave in Berlin in 1926.  Einstein questioned why Heisenberg
accepted the existence of electrons in atoms but refused to speak of their
trajectories merely because these could not be observed directly.  After all,
electron trajectories appeared to be plainly observable in cloud chambers.
Why, Einstein asked in effect, should a concept that was legitimate for an
electron in a cloud chamber suddenly become inadmissible when the electron
was inside an atom?  He was therefore dissatisfied with Heisenberg's demand
that atomic electron trajectories be excluded simply on grounds of their
non-observability \cite{Heisenberg1971}.

The discussion then turned to the more fundamental methodological issue.
When Heisenberg defended his position by appealing to Einstein's own use of
observable quantities in relativity, Einstein rejected this interpretation
of his method.  As Heisenberg later recalled, Einstein's essential reply was
that ``it is the theory which decides what we can observe''
\cite{Heisenberg1971}.  Observation is not logically prior to theory: a
theoretical framework is required to determine which physical quantities can
be inferred from experimental events and with what accuracy.

This remark made a deep impression on Heisenberg.  The relevant question was
no longer whether one should exclude electron trajectories \emph{a priori}
because they were not directly observable.  Instead, one should ask what the
new quantum mechanics itself implied about the possibility of observing or
reconstructing such trajectories.  Heisenberg was thereby led to examine
whether his theory imposed intrinsic restrictions on the simultaneous
determination of the position and momentum needed to specify a particle
trajectory.  The problem of observability was thus transformed from a
methodological postulate into a physical question to be answered by the
theory itself.

The cloud-chamber example subsequently became especially significant.
What is actually observed in a cloud chamber is not a continuous geometrical
line traced by an electron, but a sequence of localized droplets produced by
ionization.  The trajectory is inferred from these discrete events.  In
reflecting on this distinction, Heisenberg was led to ask how accurately
successive positions could be assigned to a particle and, simultaneously,
how accurately its momentum could be specified.  This line of reasoning
formed part of the conceptual route to his uncertainty analysis of 1927
\cite{Heisenberg1927}.

The episode is important for the question considered in the present paper.
Einstein's criticism did not establish that electrons necessarily possess
classical trajectories inside atoms, but neither did Heisenberg's original
observability argument establish that they do not.  It exposed instead a
distinction between three logically different questions: whether a trajectory
is directly observed, whether it can be inferred within a given theory, and
whether an underlying trajectory may exist even when it cannot be completely
reconstructed experimentally.  The uncertainty principle would place precise
limits on the second of these questions; whether those limits also settle the
third is a separate issue to which we shall return below.

\subsection{Helgoland and the birth of matrix mechanics}

In June 1925 Heisenberg went to the island of Helgoland, partly to escape a
severe attack of hay fever and partly to work without interruption. There he
made the crucial advance that led to his paper
``\"Uber quantentheoretische Umdeutung kinematischer und mechanischer
Beziehungen'' \cite{Heisenberg1925}. Instead of representing the position
of an atomic electron by a classical Fourier series for an orbit, he used
arrays of transition amplitudes indexed by pairs of stationary states. The
multiplication rule for these quantities was noncommutative.

Heisenberg himself did not initially formulate the new scheme in the
language of matrix algebra. Max Born recognized the arrays as matrices and,
with Pascual Jordan, developed the matrix formulation; Born, Heisenberg and
Jordan then produced a more systematic account of the new quantum mechanics
\cite{BornJordan1925,BornHeisenbergJordan1926}. The canonical variables
became noncommuting operators, with the characteristic relation
\begin{equation}
 [\hat q,\hat p]=i\hbar .
\label{eq:canonical_commutator_history}
\end{equation}
The classical electron orbit had disappeared from the fundamental
description. What replaced it was not a new microscopic trajectory but an
algebra of quantum observables.

\subsection{From noncommutativity to the uncertainty relation}

The explicit uncertainty principle came two years later. In 1927
Heisenberg analysed the physical content of the new kinematics and mechanics
and argued, using among other examples the celebrated gamma-ray microscope,
that attempts to determine a particle's position increasingly accurately
entail an increasing indeterminacy in its momentum \cite{Heisenberg1927}.
The now familiar statistical relation
\begin{equation}
 \Delta x\,\Delta p\geq \frac{\hbar}{2}
\label{eq:historical_uncertainty}
\end{equation}
was put into its precise form by Kennard, and subsequently generalized by
Robertson and Schr\"odinger
\cite{Kennard1927,Robertson1929,Schrodinger1930}.

This development introduced a distinction that is central to the present
paper. The mathematical inequality concerns the dispersions of observables
in a quantum state. Heisenberg's microscope supplied an intuitive
measurement-disturbance argument, while later Copenhagen discussions often
gave the ``uncertainty principle'' a broader epistemological or ontological
meaning. These claims should not automatically be identified with the
operator inequality itself.

\subsection{Mott's cloud-chamber problem: tracks without assumed orbits}

The fate of trajectories became especially vivid in the cloud chamber.
Radioactive decay is naturally represented in wave mechanics by an
approximately spherical outgoing alpha-particle wave. Yet a Wilson cloud
chamber displays narrow, approximately straight tracks, apparently the very
image of a classical particle trajectory. In 1929 Mott showed how this
particle-like phenomenon can arise within wave mechanics \cite{Mott1929}.

Mott considered the joint quantum-mechanical ionization of atoms in the
chamber. Although the emitted alpha particle is initially described by a
spherical wave, the amplitude for correlated ionizations is strongly
concentrated when the ionized atoms and the source lie approximately on a
straight line. Successive ionizations therefore form a track without the
calculation having to assume a pre-existing classical orbit. Mott's result
does not prove that an underlying trajectory is impossible. Rather, it
shows that the empirical appearance of a trajectory is not, by itself,
evidence that the standard wavefunction must be supplemented by one. This
logical distinction will become important below.

\section{Schr\"odinger's Approach: From Quantum Statistics to Matter Waves}
\label{sec:schrodinger_history}

\subsection{Bose, Einstein, and the quantum statistics of indistinguishable particles}

A second route to quantum mechanics began not with atomic transition
amplitudes but with quantum statistics. In 1924 S.~N.~Bose derived Planck's
black-body law by a new counting of photon states \cite{Bose1924}. Einstein
immediately recognized the significance of the method, translated Bose's
paper for publication, and extended the statistical reasoning from photons
to a gas of material particles \cite{Einstein1924,Einstein1925}.

Einstein's extension had a remarkable consequence. Below a critical
temperature a macroscopic population accumulates in the lowest one-particle
state---the phenomenon now called Bose--Einstein condensation. Equally
important for the development of wave mechanics, Einstein emphasized a
wave-like contribution to the fluctuations of the material gas and drew
attention to de Broglie's proposal that material particles possess an
associated wave character.

\subsection{De Broglie's matter waves}

Louis de Broglie proposed that a material particle of momentum $p$ should be
associated with a wavelength
\begin{equation}
 \lambda_{\rm dB}=\frac{h}{p},
\label{eq:debroglie_history}
\end{equation}
and developed the idea systematically in his 1924 thesis
\cite{deBroglie1925}. The proposal reversed the logic of the photon: if
radiation, classically wave-like, can display particle properties, perhaps
matter, classically particle-like, can display wave properties.

The historical situation had therefore become particularly suggestive.
Bose's new statistics, Einstein's extension of it to a material gas, and
de Broglie's hypothesis all pointed toward a wave aspect of matter.  Once
material particles were assigned a wavelength, however, an immediate and
natural question arose: \emph{what equation governs the propagation and
dynamics of such matter waves?}  De Broglie's relations specified their
wavelength and frequency, but did not yet provide a general wave equation
for a particle moving in an external field.  Finding such an equation was
the decisive next step.

It was Schr\"odinger who supplied the answer in 1926.  The historical
connection with the preceding statistical developments was one that
Schr\"odinger himself later emphasized.  As quoted by Moore, Schr\"odinger
remarked that ``wave mechanics had its birth in statistics''
\cite{MooreSchrodinger}.  The sequence from Bose through Einstein and
de Broglie to Schr\"odinger should therefore not be regarded merely as a
retrospective reconstruction.  Quantum statistics and the emerging idea of
matter waves provided the conceptual setting in which the search for a
dynamical equation for such waves became natural.  Schr\"odinger's decisive
achievement was to find that equation and to show that its stationary
solutions yielded the discrete energy levels of atomic systems.

\subsection{Schr\"odinger's wave equation}

In the first of his 1926 papers ``Quantisierung als Eigenwertproblem,''
Schr\"odinger formulated quantization as an eigenvalue problem and applied
the resulting wave equation to the hydrogen atom \cite{Schrodinger1926a}.
In modern notation the nonrelativistic equation is
\begin{equation}
 i\hbar\frac{\partial\psi}{\partial t}
 =
 \left[-\frac{\hbar^2}{2m}\nabla^2+V\right]\psi .
\label{eq:schrodinger_history}
\end{equation}
The discrete atomic energies now appeared as eigenvalues associated with
acceptable wavefunctions rather than as energies assigned to selected
classical orbits.

Schr\"odinger's original physical interpretation of $\psi$ was not Born's
later probability interpretation. He initially hoped for a more continuous
wave description of matter. Nevertheless, wave mechanics soon became the
most transparent space-time representation of nonrelativistic quantum
dynamics, while Born's probabilistic interpretation transformed the meaning
of the wave amplitude.

\subsection{The relation between wave and matrix mechanics}

The contrast with Heisenberg was initially striking. Matrix mechanics was
algebraic, discontinuous in its emphasis on transitions, and deliberately
freed from unobservable electron orbits; wave mechanics used differential
equations and continuous wavefields. Yet in 1926 Schr\"odinger showed how
the Heisenberg--Born--Jordan matrix rules could be represented within wave
mechanics and argued for the close equivalence of the two formulations
\cite{Schrodinger1926b}. In modern Hilbert-space quantum mechanics they are
understood as representations of the same underlying operator structure.

Historically, however, their conceptual routes remained different. The
equivalence of predictive formalisms does not imply that the motivating
pictures---observable transitions on one side and material waves on the
other---were identical. Nor does it settle the status of particle
trajectories. Indeed, the coexistence of these two routes makes the
trajectory question sharper: if the same quantum statistics can be expressed
through such different representations, what exactly does the uncertainty
relation tell us about an underlying microscopic motion?

\section{Quantum Uncertainty and the Question of Particle Trajectories}
\label{sec:trajectories}
\subsection{Feynman Paths and Short-Time Quantum Scaling}
\label{sec:feynman}

The Feynman path integral provides a natural starting point for
examining the relation between quantum dynamics and path geometry
\cite{Feynman1948,FeynmanHibbs1965}. Unlike Nelson's stochastic
mechanics, however, it is not a constructive stochastic theory
underlying quantum mechanics, but an alternative formulation of
quantum mechanics itself. The propagator between two space-time points
is expressed formally as a sum over paths,
\begin{equation}
K(x_f,t_f;x_i,t_i)
=
\int\mathcal D x\,
\exp\left[
\frac{i}{\hbar}S[x]
\right],
\end{equation}
where $S[x]$ is the classical action associated with the path $x(t)$.

For a free nonrelativistic particle, the short-time action over an
interval $\epsilon$ is
\begin{equation}
S_{\epsilon}
\simeq
\frac{m}{2}
\frac{(\Delta x)^2}{\epsilon}.
\label{eq:feynman_short_action}
\end{equation}
The phase in the path integral therefore changes appreciably when
\begin{equation}
\frac{m(\Delta x)^2}{\hbar\epsilon}
\sim 1,
\end{equation}
giving the characteristic short-time scaling
\begin{equation}
(\Delta x)^2
\sim
\frac{\hbar}{m}\epsilon.
\label{eq:feynman_scaling}
\end{equation}
Consequently,
\begin{equation}
|\Delta x|
\sim
\epsilon^{1/2},
\qquad
\frac{|\Delta x|}{\epsilon}
\sim
\epsilon^{-1/2}.
\label{eq:feynman_velocity_scaling}
\end{equation}
Thus the corresponding difference quotient becomes unbounded as
$\epsilon\rightarrow0$. In the standard regulated path-integral sense,
the paths contributing characteristically to the quantum propagator
exhibit Brownian-like short-distance scaling rather than the scaling of
ordinary differentiable trajectories \cite{AbbottWise1981}.

This statement requires an important qualification. The real-time
Feynman path integral does not define a classical stochastic process
with a positive probability measure over trajectories. Its weight
\begin{equation}
\exp\left(\frac{iS[x]}{\hbar}\right)
\end{equation}
is a complex amplitude. The individual paths entering the functional integral
are therefore not sample trajectories of a Wiener process and,
without additional assumptions, cannot be identified with ontic
histories followed by a particle. Their Brownian-like scaling is a property of
the path-integral representation of quantum propagation, not by itself
a stochastic microscopic dynamics.

The connection with Brownian motion becomes more direct after analytic
continuation to imaginary time,
\begin{equation}
t=-i\tau.
\label{eq:wick_feynman}
\end{equation}
For a Hamiltonian
\begin{equation}
H
=
-\frac{\hbar^2}{2m}\nabla^2+V,
\end{equation}
the Schr\"odinger equation
\begin{equation}
i\hbar\frac{\partial\psi}{\partial t}
=
H\psi
\end{equation}
is transformed into the Euclidean evolution equation
\begin{equation}
\frac{\partial\psi}{\partial\tau}
=
\frac{\hbar}{2m}\nabla^2\psi
-
\frac{1}{\hbar}V\psi.
\label{eq:euclidean_schrodinger}
\end{equation}
For a free particle this reduces to a diffusion equation with diffusion
coefficient
\begin{equation}
D=\frac{\hbar}{2m}.
\label{eq:feynman_diffusion}
\end{equation}
The corresponding Wiener process satisfies, in one spatial dimension,
\begin{equation}
\left\langle(\Delta X)^2\right\rangle
=
2D\,\Delta\tau
=
\frac{\hbar}{m}\Delta\tau,
\label{eq:wiener_variance}
\end{equation}
which displays precisely the same short-time scaling as
Eq.~\eqref{eq:feynman_scaling}.

The Feynman--Kac correspondence therefore provides a mathematically
precise bridge between Euclidean quantum propagation and Wiener
diffusion \cite{Kac1949}. This bridge should nevertheless be
distinguished from the ontological claim that a real-time quantum
particle literally executes Brownian motion. Feynman's formulation
starts from quantum amplitudes; the appearance of Wiener paths follows
after continuation to Euclidean time.

The path integral also clarifies the role of the classical principle
of stationary action. Quantum propagation is not obtained by requiring
each contributing path to satisfy
\begin{equation}
\delta S=0.
\end{equation}
Rather, all paths contribute amplitudes. In the semiclassical limit,
rapid phase cancellation suppresses contributions far from
stationary-action trajectories, whereas neighbouring paths around
\begin{equation}
\delta S=0
\end{equation}
interfere constructively. Classical stationary-action trajectories
therefore emerge through the stationary-phase mechanism rather than
being imposed on the individual quantum paths.

For the present argument, the central result is the short-time scaling
\begin{equation}
(\Delta x)^2\sim\Delta t.
\label{eq:brownian_scaling_summary}
\end{equation}
It is characteristic of Brownian-type, nowhere-differentiable path
geometry, but in the Feynman formulation it appears within an already
quantum description. 

Thus the Brownian-like scaling of Feynman paths reveals an important
kinematic analogy, but does not by itself provide a stochastic
trajectory model of microscopic motion.  This raises the natural
question whether quantum dynamics can instead be constructed from a
genuine stochastic process with well-defined sample trajectories.
Nelson's stochastic mechanics provides the first approach to this
question.

\subsection{Bohmian Mechanics and Definite Trajectories}
\label{sec:bohm}

Bohmian mechanics provides the clearest demonstration that the
statistical uncertainty relations of quantum mechanics do not, by
themselves, exclude definite particle trajectories
\cite{Bohm1952a,Bohm1952b}. Unlike the constructive stochastic
approaches considered below, Bohmian mechanics takes the
Schr\"odinger wavefunction and its evolution as given and supplements
them with an actual configuration whose motion is guided by the
wavefunction.

For a single spinless particle, write

\begin{equation}
\psi(\mathbf{x},t)
=
R(\mathbf{x},t)
\exp\left(\frac{iS(\mathbf{x},t)}{\hbar}\right).
\label{eq:bohm_polar}
\end{equation}
Substitution into the Schr\"odinger equation separates the dynamics
into a continuity equation,

\begin{equation}
\frac{\partial \rho}{\partial t}
+
\nabla\cdot
\left(
\rho\,\frac{\nabla S}{m}
\right)
=0,
\qquad
\rho=R^2,
\label{eq:bohm_continuity}
\end{equation}
and a modified Hamilton--Jacobi equation,

\begin{equation}
\frac{\partial S}{\partial t}
+
\frac{(\nabla S)^2}{2m}
+
V
+
Q
=0,
\label{eq:bohm_HJ}
\end{equation}
where

\begin{equation}
Q
=
-\frac{\hbar^2}{2m}
\frac{\nabla^2R}{R}
\label{eq:bohm_quantum_potential}
\end{equation}
is the quantum potential. The actual particle position
$\mathbf{X}(t)$ evolves according to the guidance equation

\begin{equation}
\frac{d\mathbf{X}}{dt}
=
\left.
\frac{\nabla S}{m}
\right|_{\mathbf{x}=\mathbf{X}(t)} .
\label{eq:bohm_guidance}
\end{equation}
Thus, once the initial position is specified, the theory assigns a
definite trajectory $\mathbf{X}(t)$ to the particle. The connection
with the statistical predictions of quantum mechanics is provided by
the quantum-equilibrium hypothesis
\cite{DurrGoldsteinZanghi1992}. If the configuration of an ensemble
is distributed initially according to

\begin{equation}
\rho(\mathbf{x},0)
=
|\psi(\mathbf{x},0)|^2,
\label{eq:bohm_quantum_equilibrium}
\end{equation}
then the continuity equation and the Bohmian guidance law imply
equivariance of the $|\psi|^2$ distribution,

\begin{equation}
\rho(\mathbf{x},t)
=
|\psi(\mathbf{x},t)|^2.
\label{eq:bohm_equivariance}
\end{equation}
This quantum-equilibrium distribution provides the link between the
underlying deterministic trajectories and the Born-rule statistics
of quantum mechanics. Bohmian mechanics can therefore retain definite
particle configurations and trajectories while reproducing the
statistical predictions of ordinary quantum mechanics, including the
standard uncertainty relations.

This distinction is central to the present argument. The uncertainty
relations constrain ensemble statistics and measurement outcomes;
they do not determine the ontology of the underlying motion.
Bohmian mechanics provides an explicit counterexample to the claim
that the quantum uncertainty relations logically require the absence
of particle trajectories. What changes is not the empirical content
of the uncertainty relations, but their interpretation.

The role of Bohmian mechanics in the present comparison should,
however, be distinguished from that of Nelson and Kac. Bohmian
mechanics begins with the quantum wavefunction and supplements it
with a deterministic trajectory dynamics. Nelson instead attempts to
derive Schr\"odinger dynamics from an underlying Wiener process,
whereas the Kac construction begins with finite-speed persistent
stochastic trajectories and connects them, through diffusion and
Wick rotation, respectively with nonrelativistic stochastic
kinematics and relativistic quantum amplitude dynamics. These
approaches therefore reach the question of microscopic trajectories
from fundamentally different directions.

\subsection{Decoherent Histories and Trajectories}

A different approach to trajectories was developed through the
consistent- and decoherent-histories formulations of quantum mechanics,
introduced by Griffiths and developed further by Omn\`es and by
Gell-Mann and Hartle \cite{Griffiths1984,Omnes1992,GellMannHartle1993}.
These approaches are particularly useful for closed quantum systems,
including the universe as a whole, where an external observer need not
be invoked. Probabilities can be assigned to suitably coarse-grained
alternative histories when interference between distinct histories is
sufficiently suppressed. Approximately classical trajectories can then
emerge without assuming microscopic particle trajectories. This bears
a conceptual resemblance to Mott's explanation of cloud-chamber tracks,
in which correlated ionizations favour approximately straight tracks
without postulating predetermined paths \cite{Mott1929}. Unlike Mott's
specific calculation, decoherent histories supplies a general framework
for consistent probability assignments to alternative histories.

\subsection{Constructive Stochastic Theories}

The path-integral formulation discussed in the preceding section
provides a powerful representation of quantum dynamics in terms of
alternative histories. It does not, however, require these paths to
represent actual stochastic trajectories followed by a physical
system. A different line of inquiry asks whether quantum dynamics can
instead be obtained from an underlying stochastic process with a
definite dynamical structure.

Such approaches may be called \emph{constructive stochastic theories}.
Rather than taking the quantum state and its evolution as the starting
point, they seek stochastic processes in configuration space from which
characteristic structures of quantum mechanics can be derived. The best-known example
is Nelson's stochastic mechanics, in which Schr\"odinger dynamics
emerges from a diffusion process with appropriately related forward
and backward stochastic derivatives
\cite{Nelson1966,Nelson1985}. Other stochastic constructions explore
different microscopic processes and different assumptions concerning
persistence, reversibility, and finite propagation speed.

These theories are of particular interest here because they shift the
question from how quantum amplitudes may be represented in terms of
paths to whether quantum or quantum-like dynamics can \emph{emerge from
an underlying stochastic process}. This distinction becomes central
when we compare Wiener diffusion with finite-speed persistent motion
and examine the very different microscopic path geometries associated
with them.

\subsubsection{Nelson's Stochastic Mechanics}

One of the most developed attempts to construct quantum mechanics from
an underlying stochastic dynamics is Nelson's stochastic mechanics
\cite{Nelson1966,Nelson1985}. The central idea is that a particle
undergoes a continuous Markov diffusion in configuration space.
Quantum behaviour is not imposed through a wavefunction at the outset;
rather, the Schr\"odinger equation is recovered from the kinematics and
dynamics of the stochastic process.

For a particle with position $\mathbf{x}(t)$, Nelson introduces forward
and backward stochastic differential equations,
\begin{align}
d\mathbf{x}(t)
&=
\mathbf{b}(\mathbf{x},t)\,dt+d\mathbf{W}(t),
\\
d\mathbf{x}(t)
&=
\mathbf{b}_{*}(\mathbf{x},t)\,dt+d\mathbf{W}_{*}(t).
\end{align}
Here $\mathbf{b}$ and $\mathbf{b}_{*}$ are, respectively, the forward
and backward drift velocities. The Wiener increments satisfy
\begin{equation}
\langle dW_i\,dW_j\rangle
=
2\nu\,\delta_{ij}\,dt,
\end{equation}
with diffusion coefficient $\nu$.

It is useful to introduce the current and osmotic velocities,
\begin{equation}
\mathbf{v}
=
\frac{1}{2}
\left(
\mathbf{b}+\mathbf{b}_{*}
\right),
\qquad
\mathbf{u}
=
\frac{1}{2}
\left(
\mathbf{b}-\mathbf{b}_{*}
\right).
\end{equation}
If $\rho(\mathbf{x},t)$ denotes the probability density, consistency
between the forward and backward diffusion equations gives
\begin{equation}
\mathbf{u}
=
\nu\nabla\ln\rho.
\end{equation}
The current velocity obeys the continuity equation,
\begin{equation}
\frac{\partial\rho}{\partial t}
+
\nabla\cdot(\rho\mathbf{v})
=
0.
\end{equation}
For an irrotational current flow, write
\begin{equation}
\mathbf{v}
=
\frac{1}{m}\nabla S.
\end{equation}
With the choice
\begin{equation}
\nu=\frac{\hbar}{2m},
\end{equation}
The connection between stochastic dynamics and a variational
Hamilton--Jacobi formulation was developed systematically by
Guerra and Morato \cite{GuerraMorato1983}, providing an important
variational foundation for the stochastic-mechanical construction.
The corresponding modified Hamilton--Jacobi equation may be written
as
\begin{equation}
\frac{\partial S}{\partial t}
+
\frac{(\nabla S)^2}{2m}
+
V
-
\frac{\hbar^2}{2m}
\frac{\nabla^2\sqrt{\rho}}{\sqrt{\rho}}
=
0.
\end{equation}
Together with the continuity equation, this is equivalent to the
Schr\"odinger equation upon defining
\begin{equation}
\psi(\mathbf{x},t)
=
\sqrt{\rho(\mathbf{x},t)}
\exp\left(\frac{iS(\mathbf{x},t)}{\hbar}\right).
\end{equation}
One then obtains
\begin{equation}
i\hbar\frac{\partial\psi}{\partial t}
=
\left[
-\frac{\hbar^2}{2m}\nabla^2+V
\right]\psi.
\end{equation}

This construction is conceptually important for the present argument.
The complex quantum amplitude is not introduced merely as a
calculational device for summing alternative paths. Instead,
Schr\"odinger dynamics is reconstructed from an underlying stochastic
process in ordinary configuration space, together with the additional
dynamical assumptions of Nelson's theory. In this sense, Nelson's
stochastic mechanics provides a particularly clear example of a
constructive stochastic approach to quantum mechanics.

Its microscopic kinematics is nevertheless very different from that
of an ordinary differentiable trajectory. Wiener sample paths are
continuous but almost surely nowhere differentiable. Consequently,
although the forward and backward drift velocities
$\mathbf{b}$ and $\mathbf{b}_{*}$ are well defined, they are not
ordinary instantaneous velocities obtained by differentiating an
individual sample path. This distinction will be central when we turn
to finite-speed persistent stochastic dynamics.

A frequently discussed qualification of Nelson's stochastic mechanics is
the Wallstrom objection \cite{Wallstrom1994}.  Nelson's construction
derives the Schr\"odinger dynamical equation from an underlying stochastic
mechanics, but the stochastic Hamilton--Jacobi--Madelung equations do not,
by themselves, enforce the global circulation condition
\begin{equation}
\oint \nabla S\cdot d\mathbf{x}=nh,
\qquad n\in\mathbb{Z}.
\end{equation}
The condition follows if the associated phase factor
$\exp(iS/\hbar)$ is required to be single-valued (more generally, if the
appropriate global admissibility condition on the quantum state is
imposed).  The significance of Wallstrom's observation must therefore be
stated carefully.  Conventional quantum mechanics likewise supplements
the Schr\"odinger equation with conditions specifying its admissible
physical states.  Thus the objection does not prevent Nelson's theory
from reproducing quantum mechanics once the corresponding global
condition is imposed; rather, it shows that this condition is not itself
derived from the underlying stochastic-hydrodynamic dynamics.  Its force
depends on whether stochastic mechanics is required to derive the entire
quantum formalism, including its global state-space restrictions, or the
dynamical equation together with an independently specified class of
admissible states.

Various resolutions or modifications of Nelson-type stochastic mechanics
have been proposed \cite{CarlenLoffredo1989,Schmelzer2011,Derakhshani2015}.
Related constructive approaches obtain quantum behaviour from additional
geometrical or path structure, including Ord's self-quantizing paths,
Kuipers' stochastic mechanics on complex pseudo-Riemannian manifolds, and
Nottale's scale-relativistic construction
\cite{OrdMann2003,Kuipers2022,Nottale1996,Nottale2011}.  We shall not
pursue these approaches here.  As will be seen in Section~5, the
finite-speed Kac--GJKS construction follows a different route: the quantum
amplitude equation is obtained directly from the two-component persistent
process by Wick rotation rather than through a Madelung hydrodynamic
reconstruction.  The specific Wallstrom circulation problem therefore
does not arise at this stage, although the resulting wave description,
like any quantum or quantum-like theory, must still be supplied with the
appropriate global conditions defining its admissible states.

\subsection{Finite-Speed Stochastic Dynamics: The Kac Process}

Nelson's stochastic mechanics demonstrates that Schr\"odinger dynamics
can be reconstructed from an underlying stochastic process. Its basic
stochastic kinematics is, however, Wiener diffusion. This raises a
natural question: must diffusion itself be regarded as fundamental, or
can it arise as the limiting form of a more elementary stochastic
process?

A particularly simple alternative is provided by persistent stochastic
motion in which propagation occurs at finite speed and the direction of
motion changes only at randomly distributed times. Such processes were
studied by Goldstein in connection with the telegrapher equation and
were subsequently given a particularly transparent stochastic
formulation by Kac \cite{Goldstein1951,Kac1974}. The resulting
\emph{Kac process} provides a direct bridge between finite-speed
persistent motion and ordinary diffusion.

\subsubsection{Definition of the Kac process}

Consider a particle moving on a one-dimensional line with fixed speed
$v$. Its instantaneous velocity is
\begin{equation}
\dot{x}(t)=v\,s(t),
\qquad
s(t)=\pm1,
\label{eq:kac_velocity}
\end{equation}
so that the particle moves either to the right with velocity $+v$ or
to the left with velocity $-v$.

The sign variable $s(t)$ reverses at random times. Let $N(t)$ be a
Poisson counting process with rate $\lambda$. Then
\begin{equation}
s(t)=s(0)(-1)^{N(t)},
\label{eq:kac_sign}
\end{equation}
and hence
\begin{equation}
x(t)
=
x(0)+v\int_0^t s(t')\,dt'.
\label{eq:kac_trajectory}
\end{equation}
An individual realization therefore consists of straight-line segments
with slopes $+v$ and $-v$, joined at Poisson-distributed reversal
times. The probability that no reversal occurs during an interval of
duration $t$ is
\begin{equation}
P_0(t)=e^{-\lambda t},
\end{equation}
while the probability of exactly $n$ reversals is
\begin{equation}
P_n(t)
=
e^{-\lambda t}
\frac{(\lambda t)^n}{n!}.
\end{equation}
The mean waiting time between successive reversals is therefore
\begin{equation}
\tau_{\rm flip}
=
\frac{1}{\lambda}.
\label{eq:kac_flip_time}
\end{equation}
The contrast with Wiener diffusion is immediate. A Wiener trajectory
is continuous but almost surely nowhere differentiable, whereas a Kac
trajectory is continuous and piecewise differentiable. Its velocity is
the finite quantity $+v$ or $-v$ at every time except the discrete
reversal points, where the two-sided derivative is undefined but the
one-sided velocities remain finite.

\subsubsection{Directional probabilities}

Let $P_{+}(x,t)$ denote the probability density for the particle to be
at $x$ at time $t$ while moving with velocity $+v$, and let
$P_{-}(x,t)$ denote the corresponding density for velocity $-v$.
Probability balance over an infinitesimal time interval gives
\begin{align}
\frac{\partial P_{+}}{\partial t}
&=
-v\frac{\partial P_{+}}{\partial x}
-\lambda P_{+}
+\lambda P_{-},
\label{eq:kac_plus}
\\
\frac{\partial P_{-}}{\partial t}
&=
+v\frac{\partial P_{-}}{\partial x}
+\lambda P_{+}
-\lambda P_{-}.
\label{eq:kac_minus}
\end{align}
The first term on the right-hand side of each equation describes
deterministic transport at finite speed, while the remaining terms
describe stochastic transfer between the two velocity sectors.

Define the total probability density
\begin{equation}
P=P_{+}+P_{-}
\label{eq:kac_total}
\end{equation}
and probability current
\begin{equation}
J=v(P_{+}-P_{-}).
\label{eq:kac_current}
\end{equation}
Adding Eqs.~\eqref{eq:kac_plus} and \eqref{eq:kac_minus} gives the
continuity equation
\begin{equation}
\frac{\partial P}{\partial t}
+
\frac{\partial J}{\partial x}
=
0,
\label{eq:kac_continuity}
\end{equation}
whereas subtraction gives
\begin{equation}
\frac{\partial J}{\partial t}
+
2\lambda J
=
-v^2\frac{\partial P}{\partial x}.
\label{eq:kac_current_equation}
\end{equation}
Eliminating $J$ between these equations yields
\begin{equation}
\frac{\partial^2P}{\partial t^2}
+
2\lambda\frac{\partial P}{\partial t}
=
v^2\frac{\partial^2P}{\partial x^2},
\label{eq:telegraph_equation}
\end{equation}
the telegrapher equation.

Unlike the diffusion equation, Eq.~\eqref{eq:telegraph_equation} is
hyperbolic and retains a finite characteristic propagation speed $v$.
This finite-speed structure is the central kinematic feature that is
lost in the diffusion limit.

\subsubsection{Persistence and the diffusion limit}

The persistence of the process can be expressed directly through its
velocity correlation. For the symmetric process,
\begin{equation}
\langle s(t)s(0)\rangle
=
e^{-2\lambda t},
\end{equation}
and therefore
\begin{equation}
\langle\dot{x}(t)\dot{x}(0)\rangle
=
v^2e^{-2\lambda t}.
\label{eq:kac_velocity_correlation}
\end{equation}
The velocity-correlation time is consequently
\begin{equation}
\tau_{\rm corr}
=
\frac{1}{2\lambda}.
\end{equation}
This should be distinguished from the mean waiting time between
successive reversals,
\begin{equation}
\tau_{\rm flip}
=
\frac{1}{\lambda}.
\end{equation}
The factor of two arises because each reversal changes the sign of the
velocity.

At times long compared with the persistence timescale, the current in
Eq.~\eqref{eq:kac_current_equation} relaxes rapidly compared with the
evolution of $P$. Neglecting $\partial J/\partial t$ to leading order
then gives
\begin{equation}
J
\simeq
-\frac{v^2}{2\lambda}
\frac{\partial P}{\partial x}.
\end{equation}
Defining
\begin{equation}
D
=
\frac{v^2}{2\lambda},
\label{eq:kac_D}
\end{equation}
one obtains Fick's law,
\begin{equation}
J
\simeq
-D\frac{\partial P}{\partial x},
\end{equation}
and hence
\begin{equation}
\frac{\partial P}{\partial t}
=
D\frac{\partial^2P}{\partial x^2}.
\label{eq:kac_diffusion_limit}
\end{equation}
Ordinary diffusion is therefore recovered as the long-time limiting
description of finite-speed persistent motion. More precisely, the
Brownian diffusion process itself is obtained under the Kac scaling
\begin{equation}
v\rightarrow\infty,
\qquad
\lambda\rightarrow\infty,
\qquad
\frac{v^2}{2\lambda}=D
\quad\text{fixed}.
\label{eq:kac_limit}
\end{equation}
In this limit the mean interval between reversals tends to zero while
the propagation speed diverges in such a way that the diffusion
coefficient remains finite. The piecewise-linear microscopic
trajectory then converges, in the appropriate stochastic sense, to
Brownian motion.

The conceptual relation to Nelson's construction is now apparent.
Nelson takes Wiener diffusion as the stochastic kinematics and, with
additional dynamical assumptions, reconstructs Schr\"odinger dynamics.
The Kac construction goes one kinematic level further back: Wiener
diffusion itself can arise as the singular limit of a stochastic
process possessing finite propagation speed, finite directional
persistence, and piecewise-differentiable trajectories.

If the limiting diffusion coefficient is chosen to have Nelson's value,
\begin{equation}
D
=
\frac{\hbar}{2m},
\label{eq:kac_nelson_D}
\end{equation}
then the Kac scaling requires
\begin{equation}
\frac{v^2}{2\lambda}
=
\frac{\hbar}{2m}.
\label{eq:kac_nelson_matching}
\end{equation}
This identifies the stochastic scale of the Kac diffusion limit with that 
employed in Nelson's mechanics. It establishes a kinematic connection 
between finite-speed persistent motion and the Wiener process underlying 
Nelson's construction, rather than a derivation of the full Nelson dynamics.

This leads naturally to the next question. If Schr\"odinger dynamics
can be constructed from diffusion, can finite-speed persistent
stochastic motion be associated, before the diffusion limit is taken,
with a correspondingly richer quantum-like dynamics?

\subsection{From the Kac Process to Dirac-Like Dynamics}

The Kac process introduced in the preceding section is entirely
classical: $P_{+}$ and $P_{-}$ are ordinary non-negative probability
densities, and their sum is the conserved total probability density.
Nevertheless, the underlying evolution has a suggestive two-component,
first-order structure. Unlike the diffusion equation, which is first
order in time but second order in space, the coupled Kac equations are
first order in both space and time. This makes possible a direct
structural comparison with relativistic first-order wave equations.

\subsubsection{Matrix form of the Kac equations}

Introduce the two-component probability vector
\begin{equation}
\mathbf{P}
=
\begin{pmatrix}
P_{+}\\
P_{-}
\end{pmatrix}.
\label{eq:kac_probability_vector}
\end{equation}
The coupled Kac equations can then be written as
\begin{equation}
\frac{\partial\mathbf{P}}{\partial t}
=
-v\sigma_z
\frac{\partial\mathbf{P}}{\partial x}
+
\lambda(\sigma_x-I)\mathbf{P},
\label{eq:kac_matrix_form}
\end{equation}
where
\begin{equation}
\sigma_x=
\begin{pmatrix}
0&1\\
1&0
\end{pmatrix},
\qquad
\sigma_z=
\begin{pmatrix}
1&0\\
0&-1
\end{pmatrix}.
\end{equation}
The occurrence of Pauli matrices at this stage has no quantum-mechanical
significance. They provide only a convenient matrix representation of
the two directional states of the classical stochastic process.

The common diagonal loss term may be removed algebraically by writing
\begin{equation}
\mathbf{P}(x,t)
=
e^{-\lambda t}\,
\boldsymbol{\Phi}(x,t).
\label{eq:kac_rescaling}
\end{equation}
Substitution into Eq.~\eqref{eq:kac_matrix_form} gives
\begin{equation}
\frac{\partial \boldsymbol{\Phi}}{\partial t}
=
-v\sigma_z
\frac{\partial\boldsymbol{\Phi}}{\partial x}
+\lambda\sigma_x\boldsymbol{\Phi}.
\label{eq:kac_reduced_matrix}
\end{equation}

\subsubsection{Wick rotation to the Dirac equation}

The reduced Kac equation obtained above is a real first-order evolution
equation for the two directional sectors of the persistent stochastic
process.  To make its relation to relativistic dynamics transparent, it
is useful first to formulate the problem explicitly in $1+1$
dimensions.

Let $\tau$ denote the real evolution parameter of the stochastic
process.  The corresponding Euclidean coordinates may be written as
$(x,c\tau)$, with interval

\begin{equation}
ds_E^2 = dx^2+c^2d\tau^2.
\label{eq:kac_euclidean_metric}
\end{equation}
The standard Wick rotation

\begin{equation}
\tau = it
\label{eq:kac_wick_rotation}
\end{equation}
then gives

\begin{equation}
ds_E^2
=
dx^2-c^2dt^2.
\label{eq:kac_minkowski_metric}
\end{equation}
The resulting two-component structure carries a natural indefinite metric of Lorentzian signature.
The null directions are therefore

\begin{equation}
dx=\pm c\,dt.
\label{eq:kac_lightcone}
\end{equation}
This geometrical setting is particularly natural for the Kac process.
Unlike Wiener diffusion, which has no finite propagation speed, the
persistent process consists of segments with a definite finite
velocity.  On identifying its physical limiting speed with $c$, its
two directional sectors correspond to propagation with velocities
$+c$ and $-c$.  Individual segments therefore follow the two null
directions of the $1+1$ dimensional lightcone, while sequences
containing reversals connect events lying on or within the causal
cone.

The transformation of the velocity under Wick rotation is not an
additional prescription.  It follows directly from its kinematic
definition.  If

\begin{equation}
v_E=\frac{dx}{d\tau},
\end{equation}
then Eq.~\eqref{eq:kac_wick_rotation} gives

\begin{equation}
v_E
=
\frac{dx}{i\,dt}
=
-i\,\frac{dx}{dt}.
\label{eq:kac_velocity_rotation}
\end{equation}
Hence, on identifying the Lorentzian propagation speed with $c$,

\begin{equation}
v_E\longrightarrow -ic.
\label{eq:kac_velocity_continuation}
\end{equation}
Thus the factor of $i$ associated with the velocity is a necessary
consequence of the Wick rotation of the time coordinate, rather than
an independent analytic continuation.

We now return to the reduced Kac equation, written in terms of the
Euclidean evolution parameter $\tau$,

\begin{equation}
\frac{\partial\boldsymbol{\Phi}}{\partial\tau}
=
-v_E\sigma_z
\frac{\partial\boldsymbol{\Phi}}{\partial x}
+\lambda\sigma_x\boldsymbol{\Phi}.
\label{eq:kac_reduced_euclidean}
\end{equation}
From $\tau=it$ one has

\begin{equation}
\frac{\partial}{\partial\tau}
=
-i\frac{\partial}{\partial t},
\label{eq:kac_time_derivative_rotation}
\end{equation}
while Eq.~\eqref{eq:kac_velocity_continuation} gives
$v_E=-ic$.  Substitution into
Eq.~\eqref{eq:kac_reduced_euclidean} therefore yields

\begin{equation}
-i\frac{\partial\Psi}{\partial t}
=
ic\,\sigma_z
\frac{\partial\Psi}{\partial x}
+
\lambda\sigma_x\Psi,
\label{eq:kac_after_wick}
\end{equation}
where the analytically continued two-component field has been denoted
by $\Psi$.  Multiplying by $-\hbar$ gives

\begin{equation}
i\hbar\frac{\partial\Psi}{\partial t}
=
-i\hbar c\,\sigma_z
\frac{\partial\Psi}{\partial x}
-
\hbar\lambda\,\sigma_x\Psi.
\label{eq:kac_wick_dirac}
\end{equation}
Introducing the mass scale through

\begin{equation}
\hbar\lambda=mc^2,
\label{eq:kac_mass_scale}
\end{equation}
one obtains

\begin{equation}
i\hbar\frac{\partial\Psi}{\partial t}
=
\left(
-i\hbar c\,\sigma_z\frac{\partial}{\partial x}
-
mc^2\sigma_x
\right)\Psi.
\label{eq:kac_dirac_final}
\end{equation}
This is the free Dirac equation in $1+1$ dimensions,

\begin{equation}
i\hbar\frac{\partial\Psi}{\partial t}
=
\left(
-i\hbar c\,\alpha\frac{\partial}{\partial x}
+
mc^2\beta
\right)\Psi,
\end{equation}
in the representation

\begin{equation}
\alpha=\sigma_z,
\qquad
\beta=-\sigma_x,
\label{eq:kac_dirac_matrices}
\end{equation}
for which

\begin{equation}
\alpha^2=\beta^2=I,
\qquad
\{\alpha,\beta\}=0.
\end{equation}
The sign choice $\beta=-\sigma_x$ is simply a representation
convention and has no physical significance \cite{Dirac1958}.

It is noteworthy that the Dirac velocity operator $c\alpha$ has
eigenvalues $\pm c$ \cite{Dirac1958}, preserving at the amplitude
level the two characteristic velocities of the underlying Kac
process.

This is the relativistic stochastic--quantum correspondence proposed
by Gaveau, Jacobson, Kac and Schulman (GJKS) \cite{GaveauJacobsonKacSchulman1984}. 
They explicitly noted that the analytic continuation relating the Poisson process to Dirac
propagation may be implemented by the Wick rotation
\[
t\rightarrow it,
\]
accompanied by the corresponding continuation
\[
v\rightarrow -iv.
\]
We follow this prescription here, preferring it to the alternative
GJKS prescription in which the Poisson reversal rate $\lambda$ is
continued to an imaginary value. In this way $\lambda$ retains its
original interpretation as a real stochastic reversal rate throughout
the construction, while the complex structure enters through the
Wick rotation.  Its magnitude determines the relativistic mass scale
through Eq.~\eqref{eq:kac_mass_scale}.  The two directional sectors of
the Kac process thereby become the two components of the analytically
continued Dirac amplitude.

An important advantage of this choice is that it preserves the
finite-speed characteristic structure of the underlying Kac process.
Under the Wick-rotation prescription this finite propagation
structure is carried directly into the Lorentzian $1+1$ dimensional
Dirac representation, providing the stochastic origin of its
light-cone structure.

\subsection*{Extension to $3+1$ dimensions and physical interpretation}

Although the preceding derivation has been presented in $1+1$
dimensions, where the two directional sectors of the Kac process make
the construction particularly transparent, it is not restricted to
one spatial dimension.  Gaveau, Jacobson, Kac and Schulman showed how
the stochastic construction can be extended to the $3+1$ dimensional
Dirac equation \cite{GaveauJacobsonKacSchulman1984}.  The $1+1$
dimensional case should therefore be regarded as the simplest setting
in which the relation between finite-speed persistent motion, Wick
rotation, and Dirac spinor dynamics can be displayed explicitly.

The physical significance of the analytic continuation deserves
separate consideration. If analytic continuation is regarded merely
as a mathematical device, the conventional interpretation is that
Lorentzian spacetime is the physical spacetime, while the Euclidean
stochastic construction is an auxiliary representation from which the
physical quantum dynamics is recovered. On this view, the Wick
rotation is primarily an analytical bridge between two mathematical
descriptions.

The present construction suggests that a second, more realist
interpretation is logically possible.  One may instead ask whether
the Euclidean stochastic space could represent an underlying arena of
the persistent dynamics, with Lorentzian spacetime and its complex
quantum amplitudes emerging through the geometric continuation
described above.  In this interpretation the Wick rotation would not
merely be a calculational device: it would express a relation between
an underlying stochastic geometry and the spacetime description in
which relativistic quantum dynamics is observed.

This possibility changes the nature of the foundational question.
The mathematical correspondence itself does not decide which of the
two descriptions is ontologically fundamental.  Rather, it raises the
question

\begin{equation}
\boxed{
\text{Which geometric arena, if either, represents the fundamental
physical level?}
}
\label{eq:fundamental_geometry_question}
\end{equation}
There are therefore two conceptually distinct readings of the same
formal construction:

\begin{equation}
\begin{aligned}
\text{Analytical reading:}\qquad
&\text{Euclidean stochastic space}
\;\longrightarrow\;
\text{mathematical continuation}
\\
&\hspace{2cm}\longrightarrow\;
\text{physical Lorentzian spacetime},
\\[1ex]
\text{Realist reading:}\qquad
&\text{underlying Euclidean stochastic space}
\;\longrightarrow\;
\text{geometric emergence}
\\
&\hspace{2cm}\longrightarrow\;
\text{observed Lorentzian quantum dynamics}.
\end{aligned}
\label{eq:two_geometric_readings}
\end{equation}
The second reading is an ontological hypothesis rather than a
consequence of the GJKS construction.  Nevertheless, the Wick-rotation
formulation makes its content particularly clear: the factor of $i$
is tied to the transformation between Euclidean and Lorentzian
geometry, while the corresponding transformation of the velocity
follows necessarily from $v=dx/dt$.  Thus the analytic continuation 
is not arbitrary; it has a definite geometric origin.

The extraordinary empirical success of the Dirac equation makes it
difficult to regard the geometric structure revealed by this
correspondence as merely accidental.  It strongly suggests that the
relation between finite-speed persistent dynamics, Wick rotation, and
Lorentzian relativistic quantum dynamics reflects a deeper physical
structure.

It is also here that the contrast with Nelson's stochastic mechanics
becomes especially clear.  As noted in Section~3.1, Wallstrom's point is
not that Nelson's construction fails to reproduce the Schr\"odinger
equation, but that the stochastic-hydrodynamic dynamics alone does not
derive the global circulation condition needed to restrict its solutions
to the usual quantum state space.  That restriction can be imposed through
the single-valuedness of the phase factor, just as conventional quantum
mechanics supplements its dynamical equation with admissibility conditions
on physical states.  The present Kac--GJKS construction, however, does not
pass through a Madelung hydrodynamic reconstruction: the complex
two-component amplitude is obtained directly by analytic continuation of
the finite-speed two-sector Kac dynamics.  Hence the specific Wallstrom
circulation issue does not arise at this stage, although appropriate global
conditions must still be imposed on the admissible wave states.

\subsubsection{Structural origin of the two-component dynamics}

The two-component structure has a direct stochastic origin. Its
components correspond initially to the two possible velocities,
$+v$ and $-v$, of the persistent process; they are not introduced
merely to reproduce the algebra of a spinor equation. Likewise, the
off-diagonal coupling originates in the stochastic reversal process.
In the Kac equations it transfers probability between the two
directional sectors. After analytic continuation, the corresponding
off-diagonal structure produces inter-component amplitude mixing.

The structural correspondence may therefore be summarized as
\begin{equation}
\text{two persistent directions}
\longrightarrow
\text{two-component first-order dynamics}.
\label{eq:kac_two_component_structure}
\end{equation}
Similarly,
\begin{equation}
\text{stochastic reversals}
\longrightarrow
\text{inter-component coupling}.
\label{eq:kac_sector_coupling}
\end{equation}
The finite propagation speed remains the characteristic speed of the
underlying persistent transport and is identified with $c$ in the
relativistic correspondence.

\subsubsection{Relation to the Nelson construction}

The comparison with Nelson's stochastic mechanics can now be made
without conflating the two constructions. Nelson begins with Wiener
diffusion and, with additional dynamical assumptions, reconstructs the
Schr\"odinger equation from forward and backward stochastic dynamics.
The Kac process begins at a different kinematic level, with finite-speed
persistent transport, and approaches Wiener diffusion only in the
appropriate diffusion limit.

The stochastic hierarchy may therefore be displayed schematically as
\begin{equation}
\text{Kac persistent dynamics}
\longrightarrow
\text{Wiener diffusion},
\label{eq:kac_diffusive_route}
\end{equation}
followed by
\begin{equation}
\text{Wiener diffusion}
\longrightarrow
\text{Nelson construction}
\longrightarrow
\text{Schr\"odinger dynamics}.
\label{eq:kac_nelson_route}
\end{equation}
The first arrow denotes the Kac diffusion limit, whereas the second
line involves the additional dynamical structure required in Nelson's
stochastic mechanics.

Before that diffusion limit is taken, the two directional sectors
remain explicit. Their coupled first-order evolution provides the
structural basis for the Dirac-like amplitude description, while
Wick rotation converts the real two-sector evolution into the complex
first-order amplitude dynamics of the Dirac equation, with the real
reversal rate setting the mass scale through
$\hbar\lambda=mc^2$.

The distinction is conceptually useful. Diffusion describes a regime
in which directional persistence has been lost under coarse-graining
or in the Kac limit, whereas the persistent description retains finite
propagation speed and directional structure explicitly. The
Kac--Dirac correspondence and the Kac--Nelson diffusion limit therefore
represent two different operations on the same underlying
two-directional stochastic architecture: analytic continuation leads
toward Dirac-like amplitude dynamics, while the diffusion limit leads
toward the Wiener kinematics used in Nelson's construction.

This distinction will be important in what follows because it separates
features arising from finite-speed persistence itself from those that
enter only after the passage to a complex quantum amplitude
description.

\subsection{Implications for Quantum Uncertainty and Underlying Trajectories}

The preceding examples point to an important distinction between
uncertainty in a quantum state and the possible existence of an
underlying dynamical process. The uncertainty relations constrain the
statistical distributions associated with quantum observables, but they
do not, by themselves, specify the microscopic character of whatever
dynamics may underlie those distributions.

This distinction is already apparent in Nelson's stochastic mechanics.
Individual sample paths exist, although they are almost surely nowhere
differentiable and their instantaneous velocities are not ordinary
mechanical sample-path velocities. Schr\"odinger dynamics is recovered
at the ensemble level from the forward and backward stochastic
dynamics, together with Nelson's additional dynamical assumptions.
Thus the uncertainty relations do not, by themselves, require the
absence of underlying paths; rather, they constrain the statistical
structure through which position and momentum are represented in the
quantum description.

The Kac construction makes the distinction in a different and
particularly transparent way. Before analytic continuation, every
realization of the process possesses a continuous,
piecewise-differentiable trajectory with finite instantaneous velocity
except at the discrete reversal points:
\begin{equation}
\dot{x}(t)=v\,s(t),
\qquad
s(t)=\pm1.
\label{eq:kac_underlying_velocity}
\end{equation}
The stochastic uncertainty at this level lies in the random sequence
of reversal events. Between successive reversals the motion has a
definite direction and finite speed.

The ensemble description nevertheless contains two coupled directional
sectors. As shown in the preceding section, the GJKS analytic
continuation maps the corresponding two-component first-order
structure into a Dirac amplitude equation. The resulting quantum
equation therefore has a well-defined structural and analytic
antecedent in a classical persistent process possessing finite-speed
sample trajectories.

The significance of the Kac--GJKS construction should be stated
carefully. It provides an explicit mathematical correspondence between
a finite-speed persistent stochastic process and the two-component
structure of the Dirac equation. Before Wick rotation the two
components describe directional probabilities evolving with a real
reversal rate $\lambda$. Wick rotation converts this real two-sector
evolution into a complex first-order amplitude dynamics, while the
reversal rate remains real and sets the Dirac mass scale through
$\hbar\lambda=mc^2$. What survives the continuation is therefore the
two-sector first-order architecture, its inter-component coupling, and
the finite characteristic speed that becomes the relativistic speed
$c$.

This observation is directly relevant to the interpretation of the
uncertainty principle. The relation
\begin{equation}
\Delta x\,\Delta p
\geq
\frac{\hbar}{2}
\label{eq:uncertainty_revisited}
\end{equation}
is a statement about the dispersions of the quantum observables
represented by the noncommuting operators $\hat{x}$ and $\hat{p}$.
It does not, by itself, constitute a theorem excluding every possible
underlying trajectory description. Rather, within the standard quantum
formalism it excludes a quantum state in which both position and
momentum have arbitrarily small dispersions in violation of their
operator uncertainty relation.

The examples considered above therefore suggest that three logically
distinct levels should be kept separate:
\begin{equation}
\begin{aligned}
\text{underlying dynamical histories}
&\longrightarrow
\text{statistical or amplitude description}
\\
&\longrightarrow
\text{quantum observables and uncertainty relations}.
\end{aligned}
\label{eq:three_levels}
\end{equation}
The meaning of the arrows depends on the theory and should not be
understood as representing a single universal derivation. Bohmian
mechanics, Nelson's stochastic mechanics, and the Kac--GJKS
construction illustrate different relations between trajectories,
stochastic processes, and quantum descriptions.

There is consequently no logical contradiction between quantum
uncertainty and the existence of an underlying dynamical history.
What cannot be inferred from the uncertainty relation alone is the
nature of such a history. Feynman's formulation shows that paths can
play a fundamental role in quantum dynamics without being sample
trajectories of a stochastic process or being assigned an ontic
interpretation. Where definite underlying trajectories are introduced,
they may be deterministic, as in Bohmian mechanics; stochastic and
diffusive, as in Nelson's construction; or persistent and finite-speed,
as in the real Kac process underlying the Kac--GJKS correspondence.
These possibilities have different mathematical, physical, and
conceptual status, and the uncertainty relations by themselves do not
select among them.

The more fundamental question is therefore how an underlying
dynamics with definite trajectories can give rise to quantum
mechanics. Formulated in this way, the uncertainty relation

\begin{equation}
\Delta x\,\Delta p\geq\frac{\hbar}{2}
\end{equation}
becomes a constraint that any successful underlying dynamics must
reproduce, rather than a prohibition on the existence of such
dynamics. The central issue is then to determine the structure of
that dynamics and its physical and empirical adequacy.

An interesting implication of the stochastic--quantum
correspondence concerns the cloud-chamber problem discussed
earlier. Classical Kac processes possess continuous
trajectories with directional persistence, making the
appearance of approximately straight tracks intuitively
natural. Mott's demonstration that Schr\"odinger wave
mechanics also predicts such tracks is therefore
particularly suggestive. Although the finite velocity
memory of the Kac process disappears in the diffusive
limit, Schr\"odinger dynamics supports directional
correlations that become manifest in the observed tracks.
This resemblance does not, by itself, establish that
quantum directional correlations are inherited from
classical persistence. This raises the intriguing possibility
that the trajectory-like features of quantum phenomena
reflect a deeper connection with underlying persistent
stochastic dynamics. Establishing such a connection
directly, rather than through the equivalence of the
resulting wave equations, remains an open problem.

\subsection{Discussion and Conclusions}

The uncertainty principle is often associated with the idea that
quantum mechanics precludes an underlying trajectory description.
The analysis presented here shows that this inference does not follow
from the uncertainty relations themselves. These relations constrain
the statistical distributions of quantum observables; they do not, by
themselves, determine the existence, regularity, or kinematical
properties of possible underlying paths.

The examples considered in this paper make this distinction explicit
in different ways. Feynman paths exhibit Brownian-like short-time
scaling, although in the real-time path integral they enter as
amplitude-weighted histories rather than as trajectories carrying an
ordinary positive probability measure. Nelson's stochastic mechanics
introduces genuine sample paths, but these are continuous and almost
surely nowhere differentiable. Bohmian mechanics, by contrast,
supplements the wavefunction with an actual configuration trajectory
having a well-defined velocity under the guidance dynamics. The
Goldstein--Kac process provides yet another path geometry: continuous,
piecewise-differentiable trajectories with finite propagation speed
and stochastic reversals.

These examples are not physically equivalent, nor are they presented
here as competing interpretations on an equal footing. Feynman's path
integral is a formulation of quantum mechanics, while Bohmian mechanics
supplements the quantum state with an actual configuration and guidance
law. Nelson's stochastic mechanics is instead a constructive theory in
which Schr\"odinger dynamics is reconstructed from an underlying
diffusion process with additional dynamical assumptions. The Kac
process begins at a still different kinematic level, with finite-speed
persistent stochastic motion. Their relevance to the present question
is that they exhibit markedly different path structures while being
related, in different ways, to quantum dynamics and quantum statistics.

The Kac construction is especially instructive because it connects two
different limiting or transformational routes. Under the Kac scaling
\begin{equation}
v\rightarrow\infty,
\qquad
\lambda\rightarrow\infty,
\qquad
\frac{v^2}{2\lambda}=D
\quad\text{fixed},
\end{equation}
finite-speed persistent motion converges to Wiener diffusion. If
$D=\hbar/(2m)$, the resulting diffusion has the stochastic scale used
in Nelson's mechanics. In a different direction, the two-sector
first-order structure of the Kac equations provides the architecture
which, under the GJKS analytic continuation, is mapped into a
one-dimensional Dirac amplitude equation. Under Wick rotation, 
the real two-sector stochastic evolution is
converted into complex Dirac amplitude dynamics, with the real
reversal rate $\lambda$ setting the strength of the inter-component
coupling and hence the mass scale through $\hbar\lambda=mc^2$.

These two routes should not be conflated. The diffusion limit is a
limit of the real stochastic process, whereas the Kac--Dirac relation
involves analytic continuation from a real probability dynamics to a
complex amplitude dynamics. The latter therefore establishes a
structural and analytic correspondence. What the two routes demonstrate 
together is that Brownian
and Dirac-like structures can both be related, in mathematically
different ways, to the same underlying architecture of finite-speed
persistent motion.

A related lesson follows from relativistic quantum kinematics. In
Dirac theory the instantaneous velocity operator is
\begin{equation}
\hat{\mathbf v}=c\boldsymbol{\alpha},
\end{equation}
and is not identical to $\mathbf p/m$. Instantaneous velocity,
momentum, trajectory differentiability, and statistical
position--momentum uncertainty are therefore distinct notions. In
particular, the eigenvalues $\pm c$ of a component of the Dirac
velocity operator do not imply a simultaneously sharp quantum
momentum. The zitterbewegung contribution further distinguishes the
instantaneous Dirac velocity from the mean translational velocity of a
massive particle.

The comparison developed here therefore suggests a hierarchy of
kinematical descriptions:
\begin{equation}
\begin{aligned}
\text{finite-speed persistent paths}
&\xrightarrow{\text{Kac limit}}
\text{Brownian paths},
\\[1mm]
\text{Kac two-sector structure}
&\xrightarrow{\text{analytic continuation}}
\text{Dirac-like amplitude dynamics},
\\[1mm]
\text{Wiener diffusion}
&\xrightarrow{\text{Nelson dynamics}}
\text{Schr\"odinger dynamics}.
\end{aligned}
\label{eq:conclusion_hierarchy}
\end{equation}
The arrows denote different mathematical relations and should not be
interpreted as a single chain of physical reductions.

The conclusion is therefore modest but consequential. The
Heisenberg--Kennard--Robertson uncertainty relations are fundamental
constraints on quantum statistics, but they do not uniquely determine
an underlying path geometry. In particular,
\begin{equation}
\Delta x\,\Delta p
\geq
\frac{\hbar}{2}
\end{equation}
does not, by itself, distinguish among deterministic trajectories,
diffusive stochastic paths, finite-speed persistent trajectories, or
the amplitude-weighted histories of the path integral. Nor does the
inequality, taken by itself, establish which, if any, of these
descriptions should be assigned fundamental physical status.

The central question is therefore how an underlying trajectory
dynamics can give rise to quantum mechanics. The uncertainty
relations are then statistical consequences that any successful
underlying theory must reproduce, rather than prohibitions on
definite microscopic motion. The constructions examined here show
that quantum mechanics is compatible with radically different
conceptions of microscopic motion. The Kac--Dirac
correspondence in particular points to the intriguing possibility
that relativistic quantum dynamics may emerge from an underlying
finite-speed persistent process.

\end{document}